\documentstyle[12pt]{article}
\def\hybrid{\topmargin -20pt	\oddsidemargin 0pt
	\headheight 0pt	\headsep 0pt
	\textwidth 6.25in	% A4 paper
	\textheight 9.5in	% A4 paper
	\marginparwidth .875in
	\parskip 5pt plus 1pt	\jot = 1.5ex}

\hybrid

\def\baselinestretch{1.2}

\catcode`\@=11

\def\marginnote#1{}
\newcount\hour
\newcount\minute
\newtoks\amorpm
\hour=\time\divide\hour by60
\minute=\time{\multiply\hour by60 \global\advance\minute by-\hour}
\edef\standardtime{{\ifnum\hour<12 \global\amorpm={am}%
	\else\global\amorpm={pm}\advance\hour by-12 \fi
	\ifnum\hour=0 \hour=12 \fi
	\number\hour:\ifnum\minute<10 0\fi\number\minute\the\amorpm}}
\edef\militarytime{\number\hour:\ifnum\minute<10 0\fi\number\minute}
\def\draftlabel#1{{\@bsphack\if@filesw {\let\thepage\relax
   \xdef\@gtempa{\write\@auxout{\string
      \newlabel{#1}{{\@currentlabel}{\thepage}}}}}\@gtempa
   \if@nobreak \ifvmode\nobreak\fi\fi\fi\@esphack}
	\gdef\@eqnlabel{#1}}
\def\@eqnlabel{}
\def\@vacuum{}
\def\draftmarginnote#1{\marginpar{\raggedright\scriptsize\tt#1}}

\def\draft{\oddsidemargin -.5truein
	\def\@oddfoot{\sl preliminary draft \hfil
	\rm\thepage\hfil\sl\today\quad\militarytime}
	\let\@evenfoot\@oddfoot	\overfullrule 3pt
	\let\label=\draftlabel
	\let\marginnote=\draftmarginnote
   \def\@eqnnum{(\theequation)\rlap{\kern\marginparsep\tt\@eqnlabel}%
\global\let\@eqnlabel\@vacuum}  }

\def\preprint{\twocolumn\sloppy\flushbottom\parindent 2em
	\leftmargini 2em\leftmarginv .5em\leftmarginvi .5em
	\oddsidemargin -.5in	\evensidemargin -.5in
	\columnsep .4in	\footheight 0pt
	\textwidth 10.in	\topmargin  -.4in
	\headheight 12pt \topskip .4in
	\textheight 6.9in \footskip 0pt
	\def\@oddhead{\thepage\hfil\addtocounter{page}{1}\thepage}
	\let\@evenhead\@oddhead	\def\@oddfoot{}	\def\@evenfoot{} }

\def\numberbysection{\@addtoreset{equation}{section}
	\def\theequation{\thesection.\arabic{equation}}}

\def\underline#1{\relax\ifmmode\@@underline#1\else
	$\@@underline{\hbox{#1}}$\relax\fi}

\def\titlepage{\@restonecolfalse\if@twocolumn\@restonecoltrue\onecolumn
     \else \newpage \fi \thispagestyle{empty}\c@page\z@
	\def\thefootnote{\fnsymbol{footnote}} }

\def\endtitlepage{\if@restonecol\twocolumn \else \newpage \fi
	\def\thefootnote{\arabic{footnote}}
	\setcounter{footnote}{0}}  %\c@footnote\z@ }

\catcode`@=12
\relax

\def\figcap{\section*{Figure Captions\markboth
	{FIGURECAPTIONS}{FIGURECAPTIONS}}\list
	{Figure \arabic{enumi}:\hfill}{\settowidth\labelwidth{Figure
999:}
	\leftmargin\labelwidth
	\advance\leftmargin\labelsep\usecounter{enumi}}}
 \relax
\def\tablecap{\section*{Table Captions\markboth
	{TABLECAPTIONS}{TABLECAPTIONS}}\list
	{Table \arabic{enumi}:\hfill}{\settowidth\labelwidth{Table
999:}
	\leftmargin\labelwidth
	\advance\leftmargin\labelsep\usecounter{enumi}}}
 \relax
\def\reflist{\section*{References\markboth
	{REFLIST}{REFLIST}}\list
	{[\arabic{enumi}]\hfill}{\settowidth\labelwidth{[999]}
	\leftmargin\labelwidth
	\advance\leftmargin\labelsep\usecounter{enumi}}}
 \relax
\makeatletter
\newcounter{pubctr}
\def\publist{\@ifnextchar[{\@publist}{\@@publist}}
\def\@publist[#1]{\list
	{[\arabic{pubctr}]\hfill}{\settowidth\labelwidth{[999]}
	\leftmargin\labelwidth
	\advance\leftmargin\labelsep
	\@nmbrlisttrue\def\@listctr{pubctr}
	\setcounter{pubctr}{#1}\addtocounter{pubctr}{-1}}}
\def\@@publist{\list
	{[\arabic{pubctr}]\hfill}{\settowidth\labelwidth{[999]}
	\leftmargin\labelwidth
	\advance\leftmargin\labelsep
	\@nmbrlisttrue\def\@listctr{pubctr}}}
 \relax
\makeatother
\newskip\humongous \humongous=0pt plus 1000pt minus 1000pt

\newif\ifdtup

\relax

\def\be{\begin{equation}}
\def\ee{\end{equation}}
\def\ba{\begin{eqnarray}}
\def\ea{\end{eqnarray}}

\def\del{\partial}

\def\r{\rho}
\def\a{\alpha}

\def\b{\beta}

\def\g{\gamma}
\def\G{\Gamma}
\def\d{\delta}

\def\e{\epsilon}

\def\th{\theta}

\def\m{\mu}
\def\n{\nu}

\def\Om{\Omega}
\def\l{\lambda}
\def\L{\Lambda}
\def\s{\sigma} 

\def\vphi{\varphi}

\def\cL{{\cal L}}
\def\cR{{\cal R}}

\def\bs{\bigskip}
\def\no{\noindent}

\def\qq{\qquad}

\def\IR{\relax{\rm I\kern-.18em R}}

\def \ha {{1\over 2}}

\def \ov {\over}

\def\IR{\relax{\rm I\kern-.18em R}}

\def\IR{\relax{\rm I\kern-.18em R}}

\def\cL{{\cal L}}
\def\cR{{\cal R}}

\begin{document}
%\draft

\renewcommand{\theequation}{\thesection.\arabic{equation}}
\newcommand{\beq}{\begin{equation}}
\newcommand{\eeq}[1]{\label{#1}\end{equation}}
\newcommand{\ber}{\begin{eqnarray}}
\newcommand{\eer}[1]{\label{#1}\end{eqnarray}}
\begin{titlepage}
\begin{center}

\hfill THU--96/12\\
\hfill hep--th/9602179\\

\vskip .4in

{\large \bf NON-ABELIAN DUALITY, PARAFERMIONS \\
AND SUPERSYMMETRY}

\vskip 0.5in

{\bf Konstadinos Sfetsos}
\footnote{e--mail address: SFETSOS@FYS.RUU.NL}\\
\vskip .1in

{\em Institute for Theoretical Physics, Utrecht University\\
     Princetonplein 5, TA 3508, The Netherlands}\\

\vskip .2in

\end{center}

\vskip .6in

\begin{center} {\bf ABSTRACT } \end{center}
\begin{quotation}\noindent
Non--Abelian duality in relation to supersymmetry is examined.
When the action of the isometry group on the complex structures 
is non--trivial, extended supersymmetry is realized non--locally 
after duality, using path ordered Wilson lines. 
Prototype examples considered in detail are, hyper--Kahler metrics 
with $SO(3)$ isometry and supersymmetric WZW models. 
For the latter, the natural objects in the non--local realizations 
of supersymmetry arising after duality are the classical
non--Abelian parafermions. The canonical equivalence of 
WZW models and their non--Abelian duals with respect to 
a vector subgroup is also established.

\vskip .5in

\end{quotation}
\vskip .2cm
February 1996\\
\end{titlepage}
\vfill
\eject

\def\baselinestretch{1.2}
\baselineskip 16 pt
\noindent

\section{ Introduction }
\setcounter{equation}{0}

Target space duality (T--duality) \cite{BUSCHER}
interpolates between effective field theories corresponding
to backgrounds with different spacetime and even topological properties.
Since strings propagating in T--dual backgrounds are equivalent and 
since the validity of the
corresponding effective field theories is limited, we may use T--duality 
as a way of probing truly stringy phenomena.
The latter have to be taken into account in order to resolve 
paradoxes that appear in attempts to describe various 
phenomena solely in terms of the local effective field theories.
Taking one step further this way of reasoning, 
we may view some long standing problems in fundamental Physics,
for instance in black holes Physics, as nothing but paradoxes of the
effectively field theory description, which will cease to exist once
string theoretical effects are properly taken into account.
Though this is a speculation at the moment, 
it provides the main motivation for this work.

The best ground to test these ideas is in the relation 
between duality and supersymmetry, 
in the presence of rotational isometries \cite{bakasII}.
In these cases 
non--local world--sheet effects have to be taken into account in order
for supersymmetry and Abelian T--duality to reconcile \cite{basfe}. 
Various aspects in this interplay between supersymmetry and Abelian 
T--duality have been considered in \cite{basfe,hassand,AAB,sferesto}.
This paper is a natural continuation of these works for the cases where
the duality group is a non--Abelian one \cite{osque}-\cite{KliSevII}
(for earlier work see
\cite{duearl} and references therein).

The plan of the paper is as follows: In section 2 we consider
2--dimensional bosonic $\s$-models that are invariant under the action of a 
non--Abelian group $G$ on the left.
We briefly review the canonical transformation that 
generates the dual $\s$--model in a way suitable 
for transformations of other geometrical
objects. Then we extend it to 
models with $N=1$ world--sheet supersymmetry.
For cases that admit $N=2$ or $N=4$ extended
supersymmetry we derive the transformation rules of the
corresponding complex structures. We show that, when
these belong to non--trivial representations
of a rotational subgroup of the duality group $G$, 
non--local world--sheet effects manifested with
Wilson lines, are necessary to restore extended supersymmetry at
the string theoretical level. Nevertheless, this appears to be lost
after non--Abelian duality 
from a local effective field theory point of view.
As examples, 4--dimensional Hyper--Kahler metrics with $SO(3)$ isometry
are considered in detail. The Eguchi--Hanson, Taub--NUT and 
Atiyah--Hitchin metrics are famous examples among them.
Explicit 
expressions for the three complex structures are given in general,
which could be useful for other independent applications.

In section 3 we consider the dual of a WZW model for 
a group $G$ with respect
to the vector action of a non--Abelian subgroup $H$. 
In such cases extended
supersymmetry is always realized non--locally after duality. We show 
how these realizations become natural using classical non--Abelian
parafermions of the $G/H$ coset conformal field theory. 
We also establish
the, so far lacking, canonical equivalence of these models.
As an example we consider the dual,
with respect to $SU(2)$, of the WZW model
based on $SU(2) \otimes U(1)$.

In section 4 we present 
our conclusions, and discuss feature directions of this work.

We have also written an appendix where we present in detail 
the canonical treatment of the models of section 3 following 
Dirac's method for constrained Hamiltonian systems.

\section{ Left invariant models }
\setcounter{equation}{0}

We consider classical string propagation in 
$d$-dimensional backgrounds that are invariant under the 
left action\footnote{In the language of Poisson--Lie T--duality
\cite{KliSevI}
we concentrate on cases of semi--Abelian doubles, where the coalgebra
is Abelian, or in other words to traditional non--Abelian duality. 
The reason is that there are no known non--trivial examples
of Poisson--Lie T--duality where supersymmetry enters also 
into the game. Nevertheless, we comment on section 4 on how 
Poisson--Lie T--duality may be used as a manifest 
supersymmetry restoration technique, 
in a string theoretical context.} of a group $G$ 
with dimension $dim(G)\leq d$. 
We may split the target space variables as
$X^M=\{X^\mu,X^i\}$, where $X^\m$, $\m=1,2,\dots \ ,dim(G)$ 
parametrize a group element in $G$ and 
$X^i,i=1,2,\dots, d-dim(G)$ are 
some internal coordinates which are inert under the group action. 
It will be convenient to think of them as parametrizing a 
group locally isomorphic to $U(1)^{d-dim(G)}$.
We also introduce a set of representation matrices
$\{t^A\}=\{t^a,t^i\}$, with $a=1,2,\dots, dim(G)$ and $i=1,2,\dots
d-dim(G)$, which we normalize to unity. 
The components of the left and right invariant
Maurer--Cartan 1--forms are defined as 
\be
L^A_M= -i Tr( t^A \hat g^{-1} \del_M \hat g )~ ,~~~~ 
R^A_M= - i Tr(t^A \del_M \hat g \hat g^{-1} ) =
C^{AB}(\hat g) L^B_M ~ , 
\label{LLC}
\ee
where $\hat g=g e^{i t_i X^i}$, with $g\in G$ and
$ C^{AB}(\hat g)=Tr(t^A \hat g t^B \hat g^{-1})$. When we 
specialize to the internal space, $L^i_j=R^i_j=\d^i{}_j$,
$C^{ij}=\d^{ij}$ and the corresponding structure constants are zero.
The inverses of (\ref{LLC}) will be denoted by $L^M_A$ and $R^M_A$ 
respectively.

The most general Lagrangian density which is manifestly 
invariant under the transformation $g\to \L g$, 
for some constant matrix $\L \in G$, is given by 
\be 
\cL  =   E^+_{AB} L^A_M L^B_N \del_+ X^M \del_- X^N ~ ,
\label{ELL}
\ee
where the couplings $E^+_{MN}$ can only 
depend on the $X^i$'s and thus are also invariant under the action
of the group $G$. For later use we also introduce $E^-_{AB}=E^+_{BA}$.
An equivalent expression to (\ref{ELL}) is
\be 
\cL  =   E^+_{ij} \del_+ X^i \del_- X^j + E^+_{ab} L^a_\m L^b_\n 
\del_+ X^\m \del_- X^\n + 
E^+_{ai} L^a_\m \del_+ X^\m \del_- X^i 
+ E^+_{ia} L^a_\m \del_+ X^i \del_- X^\m ~ .
\label{snfp}
\ee
The natural time coordinate on
the world--sheet is $\tau= \s^+ + \s^- $, while $\s= \s^+ - \s^-$
denotes the corresponding spatial variable.
The Poisson bracket of the variable $X^\m$ and its conjugate
momentum $P_\m$ is $\{X^\m(\s),P_\n(\s')\}= \d^\m {}_\n \d(\s-\s')$.
Since the only dependence of (\ref{snfp}) on the variables $X^\m$ is
via the combinations $L^a_\m\del_\pm X^\m$, it is convenient to know
the Poisson brackets of $L^a_\m \del_\s X^\m$ and $L^\m_a P_\m$. 
After a simple computation we find
\ba
&&\{ \del_\s X^\m L^a_\m (\s) , L^\n_b P_\n (\s') \} = 
f^a{}_{bc} L^c_\m \del_\s X^\m \d(\s-\s') + \d^a{}_b \del_\s \d(\s-\s') ~ ,
\nonumber \\
&& \{ L^\m_a P_\m (\s), L^\n_b P_\n (\s') \} 
= f_{ab}{}^c L^\n_c P_\n \d(\s-\s') ~ .
\label{LPPP}
\ea
At this point we perform the transformation  
$(X^\m,P_\n)\to (\tilde X^\m,\tilde P_\n)$ defined as 
\cite{zacdual,lozdual}
\be 
L^a_\m \del_\s X^\m  = \tilde P^a ~ , ~~~~~
L^{\m a} P_\m = \del_\s \tilde X^a - f^{ab} \tilde P_b ~ ,
\label{cantr}
\ee
where $f^{ab}\equiv f^{ab}{}_c \tilde X^c$.
One can show that it preserves the Poisson brackets (\ref{LPPP}) and hence
it is a canonical one.
The $X^i$'s remain unaffected by this transformation, 
so that $\tilde X^i=X^i$.
It is then a straightforward procedure to find 
the Lagrangian density to the dual to (\ref{snfp}) $\s$-model by 
applying the usual rules of canonical transformations in 
the Hamiltonian formalism. Here we only quote the final result:
\be
\tilde \cL = E^+_{ij} \del_+ X^i \del_- X^j + 
\left(\del_+ \tilde X^a - E^+_{ai} \del_+
X^i\right) (M_-^{-1})_{ab} 
\left(\del_- \tilde X^b + E^+_{bj} \del_- X^j \right) ~ ,
\label{duals}
\ee
with
\be
M^{ab}_-  = E^+_{ab} + f_{ab} ~ .
\label{Mab}
\ee
In addition conformal invariance requires the shift of the dilaton
\cite{BUSCHER} by $\ln\det(M_-)$.
The action (\ref{Mab}) was obtained in \cite{grnonab} in the traditional 
approach to non--Abelian duality, 
where one 
adds to (\ref{ELL}) a Lagrange multiplier term 
and introduces non--dynamical gauge fields 
which are then integrated out 
using their classical equations of motion.

The transformation (\ref{cantr}) 
was first applied to Principal
Chiral Models (PCMs); with $G=SU(2)$ in 
\cite{zacdual} and for general group in \cite{lozdual}.
In PCMs 
there is no internal space and $E^+_{ab}=\d_{ab}$. Hence, after 
non--Abelian duality with respect to the left action of the group
there are still conserved currents associated with the right 
action of the group which generate symmetries in the dual model. 
It is tempting to attribute the success of the 
canonical transformation (\ref{cantr}) to the existence of such 
conserved (local) currents.
However, this is not true since (\ref{duals}), 
which has generically no conserved (local) currents, 
correctly follows from (\ref{cantr}). Instead,
what is common in the models (\ref{ELL})
is the fact that the group action 
is entirely from the left.
As a consequence,
in the traditional approach with gauge fields,
we can completely fix a unitary gauge as $ g=1$,
by appropriately choosing the $ X^\m$'s.
In some sense (\ref{cantr}) is a straightforward generalization 
of the corresponding transformation for Abelian isometry groups 
\cite{AALcan}. 
We will see in section 3 that when the action of the isometry group is 
not entirely on the left or on the right, the analog of the
canonical transformation (\ref{cantr}) is radically different.

It is important to know how the world--sheet derivatives of the 
target space variables $\del_\pm X^M$
transform under the canonical transformation.
It is quite 
straightforward, and in fact easier than applying the canonical
transformation in all of its glory,
to show that (\ref{cantr}) and the fact that the canonical 
transformation preserves the Lorentz invariance of the 
2--dimensional $\s$-model action (\ref{ELL}), imply the dual model 
(\ref{duals}) as well as\footnote{
Details on the application of this method, for the case of Abelian
T--duality, can be found in \cite{sfeleuv}.}
\be
L^A_M\del_\pm X^M  = Q^A_\pm{}_B \del_\pm \tilde X^B ~ ,
\label{ldx}
\ee
where the matrix $Q_\pm$ is defined as
\be
(Q_\pm)^A{}_B = \left( \begin{array} {cc}
\pm (M_\pm^{-1})^{ab} & - (M_\pm^{-1})^{ac} E^\pm_{jc} \\
0 & \d^i{}_j \\
\end{array}
\right)~ ,
\label{Apm}
\ee
with $M_+^{ab}\equiv M_-^{ba}$.
Of course this transformation acts trivially on the internal 
variables $X^i$, as it should. 
Notice that $Q_\pm$ only depends on the dual model variables 
$\tilde X^M= (\tilde X^a,X^i)$. Also
$A_\pm \equiv t_a Q_\pm^a{}_B \del_\pm \tilde X^B$ can be
identified with
the on shell values of the gauge fields introduced 
in the traditional approach
to non--Abelian duality.
For later convenience, 
the inverse matrix to (\ref{Apm}) is also given:
\be
(Q^{-1}_\pm)^A{}_B = \left( \begin{array} {cc}
\pm (M_\pm)^{ab} & \pm E^\pm_{ja} \\
0 & \d^i{}_j \\
\end{array}
\right)~ . 
\label{Ainpm}
\ee
In terms of these matrices the metric 
corresponding to (\ref{duals}) can be written as
\be
\tilde G_{AB} = Q_\pm^C{}_A Q_\pm^D{}_B G_{CD}\ ,~~~~~ G_{CD} \equiv 
\ha( E^+_{CD} +  E^-_{CD}) ~ ,
\label{GAB}
\ee
where both expressions corresponding to the plus and the minus signs
give the same result for $\tilde G_{AB}$, as they should.

Let us consider the transformation (\ref{ldx}) for the plus sign.
It amounts to a non--local redefinition
of the target space variables $X^\m$ in the group element $g\in G$
associated with the isometry
\be
g= P e^{i \int^{\s^+}A_+ } ~ ,
\label{gpath}
\ee
where $P$ stands for path ordering of the exponential.
The integration is carried out for fixed $\s^-$ and 
connects a base point with $\s^+$. 
Since the equations of motions for (\ref{duals}) imply the 
vanishing of the field strength associated with $A_\pm$, the
expression (\ref{gpath}) for $g$ can be 
replaced by a similar one using
$A_-$ and integration carried out for fixed $\s^+$.
The dual background (\ref{duals}) 
is a local function of $\tilde X^M$ due to the fact that 
in the original background (\ref{ELL}) 
all group dependence was via the left--invariants $L^a_\m$. 
However, other geometrical
objects are not bound to have such a dependence. In these cases they become
non--local in the dual picture. We will shortly encounter examples
of that kind.

\vskip .1 cm
\noindent
\underline{N=1 world--sheet supersymmetry}: 
Any background can be made $N=1$ supersymmetric 
\cite{FRTO}. Thus, it is expected that a 
manifestly supersymmetric 
version of the non-abelian duality transformation exists. 
Indeed, this was found, in the traditional formalism, for the general model 
(\ref{snfp}) in \cite{Tyurin}.
In terms of a canonical transformation there is work for the supersymmetric
version of the non--linear Chiral Model on $O(4)$ and its dual
\cite{zacsusydual}. Since supersymmetry dictates the form of any
transformation compatible with it once the bosonic part is
known, it is straightforward 
to find the canonical transformation for the general supersymmetric model 
by applying the following procedure: First, one obtains the supersymmetric
version of (\ref{ldx}) by simply replacing the
bosonic fields and world--sheet derivatives by their respective 
superfields and world--sheet superderivatives
\be 
L^A_M(Z) D_\pm Z^M  = Q^A_\pm{}_B(\tilde Z) D_\pm \tilde Z^B ~ , 
\label{ldxs}
\ee
where $Z^M = X^M - i \th_+ \Psi_-^M + i \th_- \Psi_+^M - i \th_+ \th_- F^M$ 
is a generic $N=1$ superfield, with a similar expression for $\tilde Z^M$,
and $D_\pm =\mp i \del_{\th_\mp} \mp \th_\mp \del_\pm$. We have denoted
by $\Psi^M_\pm$ the world--sheet fermions and by $\th^\pm$ the two
Grassman variables. The highest component of the superfield $F^M$
is eliminated by using its equations of motion. 
It finally assumes the form
$F^M=i(\Om^+)^M_{N\L} \Psi^N_- \Psi^\L_+$.
Next we expand both
sides of (\ref{ldxs}) and read the corresponding transformation rules 
of the components.
We find that the
transformation of the bosonic part is given by (\ref{ldx}) with 
the right hand side modified by a quadratic 
term in the world--sheet fermions $\Psi^M_\pm$ 
of similar chirality as the world--sheet derivative,
\be
L^A_M\del_\pm X^M  = Q^A_\pm{}_B \del_\pm \tilde X^B  
- i \del_B Q_\pm^A{}_C \tilde \Psi^B_\pm  \tilde \Psi^C_\pm
+ {i \ov 2} f^A{}_{BC} Q_\pm^B{}_D Q_\pm^C{}_E \tilde \Psi^D_\pm  
\tilde \Psi^E_\pm ~ .
\label{ldxss}
\ee
We note that bosons in the dual model are composites of bosons and
fermions of the original model. 
This boson--fermion symphysis is a common characteristic of
duals to supersymmetric $\s$--models and
was first observed in 
\cite{zasymp} for the non--Abelian dual of the
supersymmetric extension of the Chiral Model on $O(4)$ and for Abelian
duality in \cite{hassand}.
Accordingly the redefinition of the group element $g\in G$, 
though similar to (\ref{gpath}), will also
involve the quadratic in the
fermions terms that appear in the right hand side of (\ref{ldxss}).
Nevertheless, since they can always be generated from the bosonic
first term we will
refrain, in the rest of the paper, from writing them explicitly.
The transformation of $L^A_M \Psi_\pm^M$ is similar to 
(\ref{ldx}),
\be
L^A_M \Psi_\pm^M  = Q^A_\pm{}_B \tilde \Psi_\pm^B ~ .
\label{trapsi}
\ee 
The on shell expression for the highest component of the superfield 
can be used to find the transformation of the generalized connection,
\ba
(\tilde \Om^\pm )^A_{BC} & =& 
(Q^{-1}_\pm L)^A{}_M (L^{-1} Q_\mp)^N{}_B 
(L^{-1} Q_\pm )^\L{}_C \left( (\Om^\pm )^M_{N\L} 
- \del_N L^D_\L L^M_D \right) \nonumber \\
&& +~ \del_B (Q_\pm)^D{}_C (Q^{-1}_\pm)^A{}_D ~ .
\label{omtil}
\ea
When the group $G$ is Abelian the transformations 
of the world--sheet fermions and of the connections
reduce to the corresponding ones in \cite{hassand}.\footnote{
There is an alternative 
expression to (\ref{omtil}) 
in which the right hand side depends manifestly on variables 
of the dual model only. It can be easily found using the
identity
%\be
%(\tilde \Om^\pm )^A_{BC} =
%(A^{-1}_\pm)^A{}_D (Q_\mp)^E{}_B 
%(A_\pm )^F{}_C \left( (\Om^\pm )^D_{EF} + K^D_{EF}\right)
%+ \ha (f_{BA}{}^D E^\mp_{DC} + f_{CA}{}^D E^\mp_{BD} + f_{BC}{}^D E^\mp_{AD} \right) \nonumber \\
%+ \del_B (A_\pm)^D{}_C (A^{-1}_\pm)^A{}_D ~ ,
%\label{omtilal}
%\ee
\be
(\Om^\pm)^M_{N\L} - \del_\L L^D_N L^M_D = 
L^M_A L^B_N L^C_\L \left( (\Om^\pm)^A_{BC} + \ha G^{AD} (
f_{C[D}{}^E E^\mp_{B]E} + f_{BD}{}^E E^\mp_{EC} )\right ) ~ ,
\label{omtilal}
\ee
where $(\Om^\pm)^A_{BC}$ are the connections defined using $E^+_{AB}$.
%and $K^D_{EF}= \ha G^{DC} ( f_{F[C}{}^D E^\mp_{E]D} + 
%f_{FC}{}^D E^\mp_{DF} )$.
}
Consider now a field $V_\pm^M$ that transforms under 
non-abelian duality similarly to (\ref{ldx}). Namely,
\be
L^A_M V_\pm^M  = Q^A_\pm{}_B \tilde V_\pm^B ~ .
\label{vdx}
\ee
We will call such a field a $(1_\pm,0)$ tensor,
since its transformation
under duality resembles that of a vector field under diffeomorphisms. 
In general, a $(n_+,n_-;m_+,m_-)$ tensor will have $n_\pm$ upper
and $m_\pm$ lower indices of the indicated chirality.
It is a straightforward computation to prove,
using (\ref{omtil}),(\ref{vdx}), that 
\be
\tilde D^\pm_A \tilde V_\pm^B = (L^{-1} Q_{\mp})^M{}_A (Q^{-1}_\pm L)^B{}_N 
D^\pm_M V_\pm^N ~ .
\label{tildv}
\ee
Hence, the covariant derivative of a $(1_\pm,0)$ tensor is 
a $(1_\pm,1_\mp)$ tensor. 
More generally the covariant derivative $D^+_M$
on a tensor of type $(n_+,0;m_+,0)$ will
transform it into a $(n_+,0;m_+,1_-)$ type--tensor. Similarly,
the covariant derivative $D^-_M$ on a $(0,n_-;0,m_-)$ type--tensor
will transform it into a $(0,n_-;1_+,m_-)$) type one.
The fact that the action
of covariant derivatives on tensors of the type we have indicated,
preserves their tensor character, is not a trivial statement. 
Any other combination of covariant derivatives on these or 
more general tensors produces objects that transform 
anomalously under duality.
For instance, the generalized curvature $R^+_{MNK\L}$, though
a tensor under diffeomorphisms,
it is not one under duality \cite{Tyurin,hassand,sferesto}. 
This is ultimately connected to the 
non--local nature of the duality transformation when the latter 
is viewed merely as a redefinition of the 
target space variables (cf. (\ref{gpath})).

\vskip .1 cm
\noindent
\underline{Extended world--sheet supersymmetry}: Conventionally,
extended $N=2$ supersymmetry \cite{zumino,ALFR,GHR}
requires that the background is such that an (almost) complex (hermitian) 
structure $F^\pm_{MN}$ in each sector, associated to the right and 
left-handed fermions, exists.
Similarly, $N=4$ extended supersymmetry \cite{ALFR,GHR,PNBW}
requires that, in each sector,
there exist three complex structures  $(F_I^\pm)_{MN}$, $I=1,2,3$. 
The complex structures are covariantly constant, with respect to 
the generalized connections, they are represented by
antisymmetric matrices and 
in the case of $N=4$ they obey the $SU(2)$ Clifford algebra.
If in addition they are integrable 
they also satisfy the Nijenhuis conditions, though 
these are not necessary for the existence of extended supersymmetry
\cite{nijhull}.
The above requirements put severe restrictions on the
backgrounds that admit a solution. 
For instance in the absence of torsion the metric should be Kahler
for $N=2$ and hyper--Kahler for $N=4$ \cite{ALFR}. 

In order to determine the fate of extended supersymmetry under 
non--Abelian duality it is useful to assign the complex structures to
representations
of the isometry group $G$. The simplest cases to consider are those
with
complex structures belonging to the singlet representation, thus remaining
invariant under the group action on the left. 
The most general form of such complex structures is
\be
F^\pm_{MN} = F^\pm_{AB} L^A_M L^B_N ~ ,
\label{FMN}
\ee
where $F^\pm_{AB}$ is an antisymmetric matrix independent of the
$X^\m$'s which obeys $(F^2)^A{}_B=-\d^A{}_B$. 
Its functional dependence on the internal
space variables $X^i$ is determined by demanding that 
(\ref{FMN}) is covariantly constant.
In order to find how (\ref{FMN}) transforms under non--Abelian
duality we consider, similarly to the case of Abelian duality \cite{basfe},
the 2--form $F^\pm = F^\pm_{MN} dX^M\wedge dX^N$
and its transformation properties induced by
(\ref{ldx}). The result is
\be
\tilde F^\pm_{AB} = Q_\pm^C{}_A Q_\pm^D{}_B F^\pm_{CD} ~ .
\label{tilFMN}
\ee
Hence, $F^\pm_{AB}$ transforms as a $(0,2_\pm)$ tensor under duality. 
Then, it
follows that $\tilde D^\pm_A \tilde F^\pm_{BC}=0$. 
Similarly, one verifies
that all properties of the original complex structure are 
properties of its duals as well.
In the case of $N=4$ with
three complex structures that are singlets, each one of them is of
the form (\ref{FMN}), with the corresponding 
$(F^\pm_I)_{AB}$, $I=1,2,3$ obeying
the $SU(2)$ Clifford algebra. They transform as in (\ref{tilFMN})
under 
duality and they similarly 
define a locally realized $N=4$ in the dual model.

Consider now cases where the complex structures transforms in a 
non--trivial represenation of the duality group $G$. This is impossible
if we only have $N=2$ extended supersymmetry since there should be at least
two complex structures to form a non--trivial representation. 
On the other hand, it
is well known that this implies the existence of a third one and 
thus we are led to consider the case of $N=4$ extended supersymmetry.
If the duality group is $SO(3)\simeq SU(2)$,
with structure constants $f_{IJK}=\sqrt{2} \e_{IJK}$ in
our normalization, then this implies that the 
Lie--derivative acts as
$\pounds_{R_I} F^\pm_J = f_{IJK} F^\pm_K$. 
Thus the complex structures 
$F^\pm_I$, $I=1,2,3$ transfrom in the triplet representation.
For bigger groups the same transformation is valid if we restrict to 
an appropriate rotational $SO(3)$ subgroup of $G$.
Let us introduce a singlet
under the group $G$ matrix $(\Phi^\pm_I)_{AB}$, which satisfies the
same properties as the matrix $(F^\pm_I)_{AB}$.
The form of the triplet complex structures is then
\be
(F^\pm_I)_{MN} = C^{IJ}(g) (\Phi_J^\pm)_{MN} ~ ,~~~~ 
(\Phi^\pm_J)_{MN} = (\Phi^\pm_J)_{AB} L^A_M L^B_N ~ ,
\label{FImn}
\ee
where $I,J = 1,2,3$, but note that, as always $A=1,2,\dots dim(G),\dots, d$.
In order to prove this, it is enough to notice that 
$\pounds_{R_I} C_{JK} = f_{IJL} C_{LK}$ and
$\pounds_{R_I} \Phi^\pm_K = 0$.
%\be
%\pounds_{R_I} C_{JK} = R^\m_I \del_\m C_{JK} = \e_{IJL} C_{LK} ~ .
%\label{LC}
%\ee
Consider now the 
effects of non--Abelian duality on complex structures 
of the form (\ref{FImn}).
The singlet factor $(\Phi^\pm_I)_{MN}$  remains local and transforms 
similarly to (\ref{tilFMN}). 
However, the matrix $C_{IJ}$ involves the group element 
$g\in G$ explicitly, which then 
will be given by the path ordered Wilson line (\ref{gpath}).
Hence, in the dual model
\be
(\tilde F^\pm_I)_{AB} = C^{IJ}(g) (\tilde \Phi^\pm_J)_{AB} ~ ,~~~~
(\tilde \Phi^\pm_J)_{AB} = Q_\pm^C{}_A Q_\pm^D{}_B (\Phi^\pm_J)_{CD} ~ .
\label{tiiFMN}
\ee
The complex structure as a whole is non--local precisely due to 
the attached Wilson line.
The question
is whether or not it can still be used to define an 
extended supersymmetry. 
The non--local complex structure (\ref{tiiFMN}) still satisfies the
$SU(2)$ Clifford algebra, but it is no longer 
covariantly constant. This is similar to the case of Abelian duality, 
as it was first found in \cite{basfe} and 
further elaborated in \cite{hassand}. Instead, they have to satisfy 
the general conditions for existence of non--local 
complex structures \cite{sferesto}
\be
\tilde  D^\pm_A (\tilde F^\pm_I)_{BC} \del_\mp \tilde X^A 
+ \tilde\del_\mp (\tilde F^\pm_I)_{BC} = 0~ ,
\label{conco}
\ee
where the tilded world--sheet derivative acts only on the 
non--local part of the complex structure.
Using (\ref{tiiFMN}), we find that (\ref{conco}) implies the following
equation for $\tilde \Phi^\pm_I$
\be
C_{IJ} \tilde D^\pm_A (\tilde \Phi^\pm_J)_{BC} 
+ C_{IE} f^E{}_{JD} Q_\mp^D{}_A (\tilde \Phi^\pm_J)_{BC} = 0~ .
\label{eqphi}
\ee
Then the transformations (\ref{tiiFMN}),(\ref{tildv})
imply 
\be
C_{IJ} D^\pm_M (\Phi^\pm_J)_{N\L} + C_{IA} f^A{}_{JB} L^B_M 
(\Phi^\pm_J)_{N\L} =0 ~ ,
\label{DMphi}
\ee
which is nothing but the covariantly constancy equation for the 
local complex structure (\ref{FImn}) rewritten as an equation
for $(\Phi^\pm_I)_{MN}$. 
Thus, we have proved that the original local $N=4$ breaks down to 
a local $N=1$, whereas the part corresponding to the 
extended supersymmetry gets realized non--locally. Nevertheless,
in a string setting $N=4$ remains a genuine supersymmetry.

\subsection*{ Hyper--Kahler metrics with SO(3) isometry }

In order to 
fully illustrate the previous general discussion 
it will be enough to focus on the special class 
of 4--dim hyper--Kahler metrics with $SO(3)$ symmetry. 
An additional reason is that hyper--Kahler geometry is 
an interesting subject by itself, 
especially in connection with the theory of gravitational
instantons, supersymmetric models and supergravity, and various
moduli
problems in monopole physics, string theory and elsewhere. 
The line element of 4--dim hyper--Kahler metrics with
$SO(3)$ symmetry, in the Bianchi IX formalism is given by
\be
ds^2 = f^2 (t) dt^2 + a_1^2 (t) \s_{1}^2 + a_2^2 (t) {\sigma}_{2}^2
+ a_3^2 (t) {\sigma}_{3}^2 ~ .
\label{metso3}
\ee
Here, ${\sigma}_{i}$, $i=1,2,3$ are the 
left--invariant 1--forms of $SO(3)$.\footnote{
Since the internal space
parametrized by the variable $t$ is 1-dimensional,
it will not be confusing to
use instead of upper case letters $I,J,K$, lower case ones $i,j,k$.
Also in order
to comply with standard notation in the literature and avoid
factors of $\sqrt{2}$ we will use $\s_i ={1\ov \sqrt{2}} L^i $
%-i Tr( \s^i_{Pauli} g^{-1} dg )$ 
for the left invariant Maurer--Cartan forms. 
Then also $f_{ijk}= \sqrt{2} \e_{ijk}$.}
%defined as $\S_i=-{i\ov 2} Tr(g^{-1} dg \s_i)$. 
In the parametrization of the group 
element in terms of Euler angles,
$g=e^{{i\ov 2}\phi \s_3} e^{{i \ov 2} \th \s_2} e^{{i\ov 2} \psi \s_3}$, 
they assume the form
\ba
{\sigma}_{1} & = & {1 \over 2} (\sin \theta \cos \psi d \phi  -
\sin \psi d \theta ) ~ , 
\nonumber\\
{\sigma}_{2} & = & {1 \over 2} (\sin \theta \sin \psi d \phi +
\cos \psi d \theta )~ , 
\nonumber\\
{\sigma}_{3} & = & {1 \over 2} (d \psi + \cos \theta d \phi ) ~ .
\label{si}
\ea
%The normalization has been chosen such that ${\sigma}_i \wedge {\sigma}_j 
%= {1 \over 2} {\epsilon}_{ijk} d {\sigma}_k$. 
The coordinate $t$ of the metric can always be chosen
so that
\begin{equation}
f(t) = {1 \over 2} a_1 a_2 a_3 ,
\label{ft}
\end{equation}
using a suitable reparametrization. 
It was established some time ago \cite{GIPO} that the second--order
differential equations
that provide the self--duality condition for the class of metrics
(\ref{metso3}) in the parametrization (\ref{ft}), can be
integrated once to yield the following first--order system in $t$:
\be
 {a_i^{\prime} \over a_i}  =  \ha \vec a^2 - a_i^2 
- 2 f {\l_i\ov a_i} ~ ,~~~~~~ i=1,2,3~ ,
%
%2 {b^{\prime} \over b} & = & c^2 + a^2 - 2 {\lambda}_2 ca
%- b^2 , \nonumber\\
%2 {c^{\prime} \over c} & = & a^2 + b^2 - 2 {\lambda}_{3} ab - c^2 ,
\label{lll}
\ee
where the three parameters ${\lambda}_i$ remain undetermined for the
moment.
The derivatives (denoted by prime) are taken with respect to $t$.
We essentially have two distinct categories of solutions to 
(\ref{lll}), depending on the values of the 
parameters ${\lambda}_{1}$, ${\lambda}_{2}$, ${\lambda}_{3}$. 
The first is
described by ${\lambda}_{1} = {\lambda}_{2} = {\lambda}_{3} = 0$ and 
the second by ${\lambda}_{1} = {\lambda}_{2} = {\lambda}_{3} = 1$.
The Eguchi--Hanson metric belongs to the first category and 
the Taub--NUT and the Atiyah--Hitchin metrics to the second.
These three cases provide the only non--trivial 
hyper--Kahler 4--metrics with $SO(3)$ isometry that are 
complete and non--singular \cite{GIRU}.

\vskip .1 true cm
\no
\underline{Complex structures}: It is known 
(see, for instance, \cite{GIRU}) that the complex 
structures for the Eguchi--Hanson metric are singlets under the 
the $SO(3)$ action whereas those for the Taub--NUT and the Atiyah--Hitchin
transform as a triplet. Moreover, for the Eguchi--Hanson and the Taub--NUT
metrics explicit expressions are known \cite{GIRU,BOVA}. For the 
Atiyah--Hitchin metric the complex structures are only known in
the Toda--frame formulation of the metric \cite{basfeII},
which was found using the 
fact that $\del/\del \phi$ is a manifest Killing vector field 
of (\ref{metso3}).\footnote{Any hyper--Kahler metric with a rotational
Killing symmetry can be formulated in the Toda--frame \cite{BOFI},
in which case the explicit expressions for the complex structures are known 
in general \cite{basfe}.}
Recently also, one of the complex structures of the Atiyah--Hitchin 
metric, in the parametrization (\ref{metso3}), appeared in \cite{IVRO}.
However, the result for the 
general metric (\ref{metso3}) is not known, so that we will proceed 
with its derivation.

We will prove that any hyper--Kahler metric that is 
$SO(3)$--invariant with line element
given by (\ref{metso3}) and (\ref{lll}), has three complex
structures given by
\ba
F_i = \left \{ \begin{array} {ccc}
 K_i~~ & {\rm if}~~ &  \l_1=\l_2=\l_3=0  \\
  C_{ij} K_j ~~&{\rm if}~~ & \l_1=\l_2=\l_3=1  \\
\end{array}
\right\} ~ ,
\label{comstru}
\ea
where $K_i$ is given by 
\be 
K_i = 2 e_0 \wedge e_i +  \e_{ijk} e_j \wedge e_k ~ ,
\label{ki}
\ee
with the tetrads defined as $e_0= f dt$ and $e_i = a_i \s_i$.
%and where the matrix
%$(C_{ij})$ defines the adjoint representation of $SO(3)$ 
%\be
%C_{ij}  = \ha Tr(\s_i g \s_j g^{-1})~ .
%\label{adc}
%\ee
In accordance with (\ref{FMN}),(\ref{FImn}) 
the $F_i$'s for $\l_i=0$ are singlets of $SO(3)$ whereas for $\l_i=1$
transform in the triplet representation.
In order to prove (\ref{comstru}) let us first note that clearly the
$K_i$'s obey the quaternionic algebra. Since $C_{ik}C_{jk}=\d_{ij}$ 
it is easy to verify that the $F_i$'s, in general,
obey the same algebra as well.
Then, it remains to prove that $D_\m(F_i)_{\n\r} =0$.
Since the torsion is zero, 
it suffices to show that $F_i$ is a closed 2--form
and that the associated Nijenhuis tensor vanishes.\footnote{This implies
that there exist an atlas such that one of the $F_i$'s is constant. 
The integrability of the quaternionic structure which 
would have implied
that an atlas existed such that all three $F_i$'s were constant,
requires much stronger conditions to be satisfied \cite{yano}.
Nevertheless, for the existence
of $N=4$ supersymmetry this integrability is not needed.}
A short computation using (\ref{lll}) to 
substitute for derivatives with respect to $t$, gives 
\be
dK_i = - 4 f \e_{ijk} \l_j a_k dt \wedge \s_j \wedge \s_k ~ .
\label{dki}
\ee
Thus, in the cases where $\l_i=0$, we find that indeed $F_i=K_i$ are 
closed forms. Next using the property
$ dC_{ij} = 2 C_{im} \e_{mjk} \s_k$ we compute that
\ba
d(C K)_i &  = & - 4 f C_{ij} \e_{jmk} (\l_m-1)
a_k dt \wedge \s_m \wedge \s_k \nonumber \\
& & +  2 C_{ij} \e_{jmk} \e_{mln} a_l a_n \s_k \wedge \s_l \wedge \s_n ~ .
\label{dfi}
\ea
It can be easily seen that the second line in the
above equation vanishes identically. 
Hence also for the cases $\l_i=1$, $F_i=C_{ij} K_j$ are closed 
froms.
Verifying the vanishing of the Nijenhuis tensor 
is a bit harder task, but nevertheless straightforward, and will
not yield any details.
%\be
%d(C K)_i = - 4 f C_{ij} \e_{jmk} (\l_m-1)
%a_k dt \wedge \s_m \wedge \s_k ~ ,
%\label{dfi}
%\ee
%and hence also for the cases $\l_i=1$, $F_i=C_{ij} K_j$ are closed 
%froms.
%Verifying the vanishing of the Nijenhuis tensor 
%defined, for each complex structure separately, as
%\be
%(N_i)^\m_{\n\r} = (F_i)^{\l}{}_\n D_{[\r} (F_i)^\m{}_{\l]} ~~ - ~~
%\left( \n \leftrightarrow \r \right)~ ,
%\label{Ni}
%\ee
%is a bit harder task, but nevertheless straightforward, and will
%not give any details.

\vskip .1 true cm
\no
\underline{The dual $\s$-model}: 
Non--Abelian duality on (\ref{metso3}) with respect to the $SO(3)$ isometry
group corresponds to a canonical transformation which for the 
world--sheet derivatives assumes the form (cf. (\ref{ldx}))
\be 
\s^\pm_i = \pm 2 e^{-\tilde\Phi} \left( {4f^2\ov a_i^2} \del_\pm \chi^i 
+ \chi^i \chi\cdot \del_\pm \chi \pm \e_{ijk} \chi_k a_k^2 \del_\pm 
\chi^j \right) ~ ,
\label{spmi}
\ee
where $\s^\pm_i$ are the $(1,0)$ and $(0,1)$ components of the
decomposition of the 1--forms (\ref{si}) on the world--sheet 
and the $\chi^i$'s represent the three variables dual 
to the Euler angles.
The dual to the background (\ref{metso3}) 
can be obtained by specializing (\ref{duals}) in this case. The 
explicit form for the fields is \cite{ALY}
\ba
&& d\tilde s^2  =  f^2 dt^2 + e^{-\tilde\Phi} ( \chi_i \chi_j + \d_{ij}
{4 f^2\ov  a_i^2}  ) d\chi_i d\chi_j ~ ,
\nonumber \\
&& \tilde B_{ij}  = - e^{-\tilde \Phi} \e_{ijk} \chi_k a_k^2 ~ ,
\nonumber \\
&& e^{\tilde \Phi}  =  4 ( 4 f^2 + a_i^2 \chi_i^2) ~ .
\label{dsbp}
\ea

\vskip .1 true cm
\no
\underline{The dual complex structures}: The dual to the 
2--form (\ref{ki}) can be obtained from (\ref{tiiFMN}) or by directly
transforming it using (\ref{spmi}). The result is
\ba
\tilde K^\pm_i  =  e^{-\tilde \Phi} \Bigg( 
\pm 4 f dt \wedge \Big( {4 f^2\ov a_i^2} d\chi^i
+ \chi^i \chi\cdot d\chi \pm \e_{ijk} \chi_k a_k^2 d\chi^j \Big) 
\nonumber \\
\pm ~ {4 f\ov a_i} \chi\cdot d\chi \wedge d\chi^i 
+ a_i^2 \e_{ijk} a_j a_k d\chi^j \wedge d\chi^k 
 \Bigg )  ~ .
\label{tif3}
\ea
For the cases where the original hyper--Kahler metric corresponds
to the choice $\l_i=0$ in (\ref{lll}) these are in fact the 
three complex structures for the dual background (\ref{dsbp}),
which has locally realized $N=4$ supersymmetry. 
It can be shown that the (anti)self--duality conditions 
of the dilaton--axion field are solved and therefore we have 
found that (\ref{dsbp}) is a new class of axionic--instantons which
are related to hyper--Kahler metrics (\ref{metso3}) via non--Abelian
duality. Though not obvious, it can be shown that
the metric in (\ref{dsbp}) is 
conformally flat
(for the case where (\ref{metso3}) is the Eguchi--Hanson metric
this was observed in \cite{ALY}), and the conformal 
factor $e^{-\tilde\Phi}$ 
satisfies the Laplace equation adapted to the flat metric.
This is in agreement with a theorem 
proved in \cite{CHS} for 4-dim backgrounds with $N=4$ world--sheet 
supersymmetry and non--vanishing torsion. The particular form of the
coordinate change needed to explicitly demonstrate this
is complicated and not very illuminating.
Here we mention the result for the non--Abelian dual 
to 4-dimensional flat space
which corresponds to the choice $a_1=a_2=a_3=(-t)^{-1/2}$ in (\ref{metso3}).
We found that the dual metric can be written in terms of
Cartesian coordinates $x_i$, 
as $d\tilde s^2=e^{-\tilde \Phi} dx_i dx_i$,
where
$e^{\tilde \Phi} = 2 r \sqrt{ r +x_4}$, with $r^2=x_i x_i$.

For the cases where the original hyper--Kahler metric corresponds
to the choice $\l_i=1$ in (\ref{lll}) the dual
background has non--locally realized $N=4$ world--sheet supersymmetry.
The complex structures are 
$\tilde F^\pm_i = C_{ij}(g) \tilde K^\pm_j $,
with $C_{ij}(g)$ being
non--local
functionals of the dual space variables according to (\ref{spmi}).

%%%%%%%%%%%%%%%%%%

\section{ Duals of WZW Models } 
\setcounter{equation}{0}

We would
like to make contact with exact 
conformal field theoretical results. The hyper--Kahler metrics
and their non--Abelian duals we have examined are not 
appropriate for such an investigation since their description
in terms of exact conformal field theories is, at present,
unknown.
The best examples to consider in this respect 
are non--Abelian duals of WZW models,
since, as it turns out, 
the non--local realizations of supersymmetry that
arise after duality can be naturally expressed 
in terms of non-abelian parafermions.

The WZW model action, to be denoted by $I_{wzw}(g)$, 
for a group element $g\in G$ corresponds to a background
with metric and torsion given by
\be
G_{MN} = L^A_M L^A_N = R^A_M R^A_N~ ,~~~~~ H_{MN\L} = f_{ABC} L^A_M
L^B_N L^C_\L = f_{ABC} R^A_M R^B_N R^C_\L ~ .
\label{wzwgb}
\ee
A WZW model for a general group can be made $N=1$ supersymmetric 
on the world--sheet \cite{veckni}. If the group is an
even dimensional one the supersymmetry is 
promoted to an $N=2$ \cite{spindel}. Moreover, WZW models 
based on quaternionic groups have actually $N=4$ \cite{spindel}. 
The general form of the complex structures is
very similar to (\ref{FMN}), 
\be
F^+_{MN} = F^+_{AB} L^A_M L^B_N ~ , ~~~~~
F^-_{MN} = F^-_{AB} R^A_M R^B_N ~ ,
\label{FLR}
\ee
where the {\it constant} matrices $F^\pm_{AB}$ are 
Lie algebra complex structures \cite{spindel}. The covariant
constancy of $F^\pm_{MN}$ follows trivially from the fact that 
$D^+_M L^A_N = D^-_M R^A_N =0 $, which are valid for any WZW model. 
It is obvious that $F^+$ ($F^-$) is invariant under the 
left (right) group action. Thus, under the
vector action 
of a non--Abelian subgroup $H$ of $G$, i.e.
$g\to \L^{-1} g \L$, none of 
the $F^+$, $F^-$ is invariant.

The analog of the canonical transformation (\ref{cantr}) 
or (\ref{ldx}) for the non--Abelian dual of a WZW model with respect
to its vector subgroup $H$ will be 
presented in the next subsection. Here, we proceed traditionally
by starting with the usual gauged WZW action \cite{gwzwall,SCH}
plus a Langrange multiplier term,
\be
S=I_{wzw}(g) + {k\ov \pi} \int 
Tr\Big( A_+ \del_- g g^{-1} -  g^{-1}\del_+ g A_-
+ A_+ g A_- g^{-1} - A_+ A_-\Big) + i Tr\Big(v F_{+-}\Big) ~ ,
\label{GWZW}
\ee
where
$A_\pm$ are gauge fields in the Lie algebra of a subgroup $H$ of $G$
with corresponding field strength 
$F_{+-}= \del_+ A_- - \del_- A_+ - [A_+,A_-]$ and $v$ are
some Lie algebra variables in $H$ that play the role of
Lagrange multipliers. We also split indices as $A=(a,\a)$,
where $a\in H$ and $\a\in G/H$.
Variation of (\ref{GWZW}) with respect to all
fields gives the classical equations of motion
\ba
\d A_+ & :& ~~~~ \qq D_- g g^{-1} \big |_H  + i D_- v =0 ~ , \label{dap}
%\nonumber 
\\
\d A_- & : & ~~~~ \qq g^{-1} D_+ g  \big |_H + i D_+ v =0 ~ , \label{dam}
%\nonumber 
\\
\d g & : & ~~~~ \qq D_+(D_- g g^{-1}) + F_{+-}=0 ~ , \label{dg}
%\nonumber 
\\
\d v & : &  \qq ~~~~ F_{+-} = 0 ~ .
\label{equmot}
\ea
To find the dual $\s$-model a unitary gauge should be chosen. 
This is done by fixing $dim(H)$ 
variables among the total number of $dim(G) + dim(H)$ ones,
thus remaining
with a total of $dim(G)$ variables, which we will denote by $X^M$.
If $H\neq G$ then generically there is no isotropy subgroup and we
can gauge fix all $dim(H)$ variables in the group element $g\in G$.
If $H=G$ then the non--trivial isotropy subgroup corresponding to the
Cartan subalgebra of $G$ cannot be gauge fixed away. In such
cases we gauge fix $dim(G)-rank(G)$ parameters in $g$ and the
remaining $rank(G)$ ones
among the Lagrange multipliers $v^a$.
Then we eliminate the gauge fields 
using their classical equations of motion (\ref{dap}),
(\ref{dam}),
\ba
A_+^a & = &+ i \big(C^T - I- f\big)^{-1}_{ab} 
\big(L^b_\m \del_+ X^\m + \del_+ v^b\big)~ \equiv ~ A_+^a{}_M \del_+ X^M ~ ,
\nonumber \\
A_-^a & = & -i \big(C - I+ f\big)^{-1}_{ab} 
\big(R^b_\m \del_- X^\m + \del_- v^b\big) ~ \equiv ~ A_-^a{}_M \del_- X^M ~ .
\label{apam}
\ea 
Finally, the dual $\s$-model is given by \cite{grnonab,sfedual}
\be 
S= I_{wzw}(g) - {k\ov \pi} \int \big(L^a_\m \del_+ X^\m + \del_+ v^a\big)
\big(C - I + f\big)^{-1}_{ab} \big(R^b_\n \del_- X^\n + \del_- v^b\big)~ .
\label{dualsmo}
\ee
A dilaton $\Phi=\ln \det\big(C-I +f\big)$ is also induced in order 
to preserve conformal invariance at 1--loop \cite{BUSCHER}.

As in the previous section it will be convenient to have an explicit 
expression for the generalized connections of the 
dual model (\ref{dualsmo}). For this we utilize the 
classical string equations for the dual action (\ref{dualsmo}),
$D_+(D_- g g^{-1})=0$, which follow from 
(\ref{dg}) after we use (\ref{equmot}). In these equations the gauged 
fields entering the covariant derivatives should be replaced by
their on shell values (\ref{apam}). We define 
\be
Tr(t^A g^{-1} D_+ g) =i \cL^A_M \del_+ X^M ~ , ~~~~~
Tr(t^A  D_- g g^{-1}) =i \cR^A_M \del_- X^M ~ .
\label{callr}
\ee
Under gauge transformations $\cL^A_M$ and $\cR^A_M$ are 
left and right invariant respectively.
Then it is easy to cast the classical equations of motion into the 
standard form for any 2--dimensional $\s$--model
\be
\del_+ \del_- X^M + (\Om^-)^M_{N\L} \del_+ X^N \del_- X^\L = 0 ~  ,
\label{class}
\ee
from which we read off the generalized connection of the 
dual model 
\be
(\Om^-)^M_{N\L} = \cL^M_A \del_\L \cL^A_N + i f^A{}_{Bc} 
\cL^M_A \cL^B_N A_-^c{}_\L ~ .
\label{omoo}
\ee
It is
convenient to define the following gauge invariant elements,
in the Lie algebra of $G$
\be
\Psi_+  = -i h_-^{-1} g^{-1} D_+ g h_- ~ ,~~~~~~~~
\Psi_-  =  -i h_+^{-1} D_- g g^{-1} h_+  ~ ,
\label{paraf}
\ee
where the group elements $h_\pm \in H$ are given
by path ordered exponentials similar to (\ref{gpath})
\be
h_+^{-1} = P e^{- \int^{\s^+} A_+}~ , ~~~~~
h_-^{-1} = P e^{- \int^{\s^-} A_-}~ , 
\label{hphm}
\ee
with the gauge fields $A_\pm$ determined by (\ref{apam}).
They obey $A_{\pm} = \del_\pm h_\pm h_\pm^{-1}$.
Using the
classical equations of motion (\ref{dap})-(\ref{equmot}), it
can be shown that $\Psi_+$ and $\Psi_-$ are chiral
\be
\del_- \Psi_+ = 0 ~ , ~~~~~~ \del_+ \Psi_- = 0 ~ .
\label{chipsi}
\ee
We will also denote $\Psi^A = \Psi^A_\pm{}_M \del_\pm X^M $, where
\be
\Psi^A_+{}_M = 
C^{BA}(h_-) \cL^B_M ~ ,~~~~~~
\Psi^A_-{}_M = 
C^{BA}(h_+) \cR^B_M ~ .
\label{chllrr}
\ee
Because they have Wilson lines attached to them, $\Psi_\pm$ are 
non--local. Since, the action we started with (\ref{GWZW}) 
contains the standard gauged WZW action corresponding to the
coset $G/H$,
it is expected that $\Psi_\pm$ will be related to the
classical non--Abelian parafermions \cite{BCR,BCH}. The precise 
relationship will be uncovered in the next subsection.

We are now in the
position to examine the fate of world--sheet supersymmetry
under non--Abelian duality. 
We will show that the dual action (\ref{dualsmo}) has
non--locally realized extended supersymmetry with complex structures,
corresponding to (\ref{FLR}), given by
\be
\tilde F^+_{MN} = F^+_{AB} \Psi_+^A{}_M \Psi_+^B{}_N ~ , ~~~~~~~
\tilde F^-_{MN} = F^-_{AB} \Psi_-^A{}_M \Psi_-^B{}_N ~ .
\label{FLRpsi}
\ee
It is obvious that the dual complex structures (\ref{FLRpsi}) 
obey all properties of their counterparts (\ref{FLR})
except that they are not covariantly constant. 
Being non--local they 
should satisfy instead, the equation \cite{sferesto}
\be
\tilde  D^\pm_M (\tilde F^\pm)_{N\L} \del_\mp \tilde X^M 
+ \tilde\del_\mp (\tilde F^\pm)_{N\L} = 0~ ,
\label{conpsi}
\ee
where the tilded derivative acts only on the non--local part of the 
complex structures contained in $h_\pm$, which are given by
the path ordered exponentials (\ref{hphm}).
For this it is enough to prove that 
\be
\tilde D^\pm_M \Psi^A_{\pm} \del_\mp X^M 
+ \tilde \del_\mp \Psi^A_\pm = 0 ~ ,
\label{dpsidpsi}
\ee
where, similarly to (\ref{conpsi}),
the tilded world--sheet derivative acts only on the
non--local part of $\Psi^A_\pm$.
This becomes a straightforward computation after we 
use the expression for the generalized connection of the 
dual model (\ref{omoo}). 

Thus, we have shown that as long as $H$ 
is non--Abelian, T--duality breaks all local 
extended supersymmetries which are then realized non--locally 
with complex structures given by (\ref{FLRpsi}). 
Our treatment is equally applicable to
the cases where $H$ is an Abelian subgroup of $G$. 
However, in such cases 
T--duality preserves one extended supersymmetry.
In order to see
that let us recall \cite{spindel} 
that for any even dimensional WZW model the 
non--vanishing elements of the matrix $F^\pm_{AB}$ in the
Cartan basis are $F^\pm_{\a \bar \a}=i$ and $F^\pm_{ij}$, where 
$i,j$ here are labels in the Cartan subalgebra of $G$
and $\a$ ($\bar \a$) is a positive (negative) root label. 
Since the group $H$ is Abelian we have $C_{ij}(h_\pm) =\d_{ij}$.
Using the fact that
$C_\b{}^\a{}(h_\pm) C_{\bar \g}{}^{\bar\a}(h_\pm) = \d_{\b\bar\g}$
and (\ref{chllrr}), we find that the complex 
structures (\ref{FLRpsi}) 
are local functions of the target space variables
and assume the form
\ba
\tilde F^+_{MN} & = & i \cL^\a_{[M} \cL^{\bar \a}_{N]}
+ F^+_{ij} \cL^i_M \cL^j_N ~ ,
\nonumber \\
\tilde F^-_{MN} & = & i \cR^\a_{[M} \cR^{\bar \a}_{N]}
+ F^-_{ij} \cR^i_M \cR^j_N ~ .
\label{abco}
\ea
We conclude that, if $H$ is Abelian
T--duality preserves the local $N=2$
of the even dimensional supersymmetric WZW models. 
However, this is not the case for the two additional
complex structures present in WZW models based on quaternionic 
groups, which actually have $N=4$ extended supersymmetry. 
These cannot be written in a form similar to (\ref{abco}) and 
remain genuinely non--local. More details for the case of the
WZW model based on $SU(2) \otimes U(1)$ can be found in 
\cite{basfe,basfeII,kerk95} and for a general quaternionic group
in \cite{sferesto}.

\subsection*{Non--Abelian parafermions}

We will now find the precise
relation of $\Psi_\pm$ to the non--Abelian classical
parafermions of the coset theory $G/H$ \cite{BCH}. 
Moreover, we will show that their Poisson brackets 
obey the same algebra as the currents of the
original WZW model.
This provides the, so far
lacking, canonical equivalence between a WZW model for $G$ and its 
dual with respect to a vector subgroup $H$ 
as it is given by (\ref{dualsmo}). 
In retrospect the emergence 
of parafermions is not a surprise since the non--Abelian
duals of WZW models are related to gauged WZW models, 
as it was shown in \cite{sfedual} and \cite{ALY}.

Since we are interesting in the computation of Poisson brackets,
our treatment here will be completely classical. Hence,
the non--trivial Jacobians arising from changing
variables inside the functional path 
integral \cite{SCH} will be ignored.
Let us define the gauge invariant analogs of $g,h,v$ as
\be
f= h_-^{-1} g h_- \in G , ~~~~~ h= h_+^{-1} h_- \in H ~ , ~~~~~ 
\tilde v = h_-^{-1} v h_- \in H ~ .
\label{fhv}
\ee
and introduce a group element $\l\in H$ such that the 
$i \del_- \tilde v = - \del_- \l \l^{-1}$.
With these definitions the gauge field strength 
$F_{+-}= h_- \del_- (h^{-1}\del_+ h) h^{-1}_- $.
Then with the
help of the Polyakov--Wiegman formula the action (\ref{GWZW})
assumes the form
\be
S= I_{wzw}(h f) - I_{wzw}(h \l) + I_{wzw}(\l) ~ .
\label{sss}
\ee
The form of 
$\Psi_-$, defined in (\ref{paraf}), in terms of gauge invariant 
quantities is $\Psi_-= -i h \del_- f f^{-1} h^{-1}$. 
The latter expression contains $h_+$ whose definition 
(\ref{hphm}) involves a timelike integral,
when we regard $\s^+$ as ``time''. This makes 
the computation of the corresponding Poisson brackets
very difficult to perform. Thus,
as in \cite{BCR,BCH}, we make use of the equation of motion $F_{+-}=0$
to replace $h_+$ by $h_-$  in the definition of $\Psi_-$ in (\ref{paraf})
or equivalently to consider Poisson brackets of \footnote{
From now on we concentrate on one chiral sector only.
We will use $x$ or $y$ to denote the world--sheet coordinate $\s^-$.}
\be
\Psi = {i k\ov \pi} \del_- f f^{-1} ~ ,
\label{mofpsi}
\ee
where for notational convenience we have modified
the normalization factor 
and have dropped the minus sign as a subscript.
We should point out that the on shell condition $\del_+ \Psi =0$ is still
obeyed. The computation of the Poisson brackets using directly the
action (\ref{sss}) will be done systematically in the appendix 
using Dirac's canonical approach to constrained 
systems. Here we follow a shortcut which enables us to make direct contact
with the parafermions. We rewrite the action (\ref{sss})
by shifting $h\to h\l^{-1}$, as \cite{ALY}
\be 
S= I_{wzw}(h \l^{-1} f) - I_{wzw}(h) + I_{wzw}(\l) ~ .
\label{sssf}
\ee
The first two terms correspond to the gauged WZW action for the
coset $G/H$ and the third to an additional WZW action.
Parafermions are introduced, similarly 
to \cite{BCR,BCH}, by defining
\be
\Psi^{G/H} = {i k\ov \pi} \del_- (\l^{-1} f) f^{-1} \l  ~ ,
\label{psigh}
\ee
where the superscript emphasizes that they are valued in the
coset $G/H$.
Their Poisson brackets have been computed in \cite{BCR,BCH} 
\ba
\{ \Psi^{G/H}_\a(x),\Psi^{G/H}_\b(y)\} & = & - {k\ov \pi} \d_{\a\b}
\d^\prime (x-y)
- f_{\a\b\g} \Psi^{G/H}_\g (y) \d(x-y)
\nonumber \\
& & - {\pi\ov 2k}
f_{c\a\g} f_{c\b\d}~ \e(x-y) \Psi^{G/H}_\g(x) \Psi^{G/H}_\d (y) ~ ,
\label{poipar}
\ea
where the antisymmetric step function 
$\e(x-y)$ equals $+1 (-1)$ if $x>y$ ($x<y$). 
The last term in (\ref{poipar}) is responsible for their non--trivial
monodromy properties and unusual statistics.
The currents corresponding to the WZW  model 
action $I_{wzw}(\l)$ in (\ref{sssf})
are defined as 
\be
J ={ i k\ov \pi}  \del_-\l \l^{-1} = { k\ov \pi} \del_- \tilde v~ ,
\label{ja}
\ee
with $\del_+ J =0$ on shell.
Using the basic Poisson bracket for a WZW model \cite{Wwzw}
\be
\{ Tr(t^a \l^{-1} \d \l)(x),Tr(t^b \l^{-1} \d \l)(y) \}= 
-{\pi \ov 2 k} \e(x-y) \d^{ab}  ~ ,
\label{bssi}
\ee
and the variation under infinitesimal transformations
\be
\d J_a = {i k\ov \pi} C_{ab}(\l) 
Tr\Big(t^b \del_- (\l^{-1} \d \l)\Big) ~ ,
\label{dja}
\ee
one proves that the following current algebra is obeyed \cite{Wwzw}
\be
\{ J_a(x),J_b (y)\}  = - {k\ov \pi} \d_{ab} \d^\prime (x-y)
- f_{abc} J_c (y) \d(x-y) ~ .
\label{curpar}
\ee
In addition due to the ``decoupling'' in (\ref{sssf})
we have $\{ \Psi^{G/H}_\a , J_b \} = 0$.
In order to compute the 
Poisson brackets of (\ref{mofpsi}) we first note that $\Psi_a = J_a$,
due to (\ref{dap}). Hence,
the bracket $\{\Psi_a,\Psi_b\} $ is the same as (\ref{curpar}). On the 
other hand $\Psi_\a= C_{\a\b}(\l) \Psi^{G/H}_\b$. To determine
$\{\Psi_\a, \Psi_\b\}$ and $\{\Psi_\a, J_b\}$ we need the variation 
\be
\d \Psi_\a = C_{\a\b}(\l) \d \Psi^{G/H}_\b + i Tr(t^b \l^{-1} \d \l)
 f_{b\g\d} C_{\a\d}(\l) \Psi^{G/H}_\g ~ .
\label{varpsi}
\ee
Then, using (\ref{poipar}), (\ref{bssi}) and (\ref{varpsi}) we find
\be
\{ \Psi_\a(x),\Psi_\b(y) \}= - {k\ov \pi} \d_{\a\b} \d^\prime (x-y)
- \Big( f_{\a\b\g} \Psi_\g (y) + f_{\a\b c} J_c (y)\Big) \d(x-y) ~ ,
\label{psipa}
\ee
and 
\be
\{ J_a(x),\Psi_\b (y) \}  = - f_{a \b \g} \Psi_\g (y) \d(x-y) ~ .
\label{siaahj}
\ee
Thus the closed algebra obeyed by $\Psi_A=\{J_a,\Psi_\a\}$
is given by (\ref{curpar}), (\ref{psipa}) and (\ref{siaahj}),
which is the current algebra for $G$. We
emphasize the fact that, even though the $\Psi_\a$'s are related to the 
coset parafermions $\Psi^{G/H}_\a$'s, they are not parafermions
themselves since in their Poisson bracket (\ref{psipa}) there is 
no term similar to the third term in (\ref{poipar}). The reason, is
precisely the ``dressing'' provided by the extra 
fields (Lagrange multipliers). This is equivalent to the 
well known realizations of current algebras in conformal field theory
using parafermions.
Hence, we have shown a canonical equivalence between a WZW model for
a general group $G$ and its dual with respect to a vector subgroup $H$
in the sense that the algebras obeyed by the natural 
(equivalently, symmetry generating) objects in the
two models are the same.

\subsection*{Non--Abelian dual to $SU(2) \otimes U(1)$}

The corresponding WZW action is 
given by
\be 
S = {k\ov 4\pi} \int \del_+ \phi \del_- \phi + \del_+ \th \del_- \th
+ \del_+ \psi \del_- \psi + 2 \cos\th \del_+ \phi \del_- \psi + 
\del_+\r \del_- \r ~ .
\label{su2}
\ee
This is the most elementary non--trivial model with
$N=4$ world--sheet supersymmetry.
The three complex structures in the right sector are given by
\be
F^+_i = 2 d\rho \wedge \s_i -  \e_{ijk} \s_j \wedge \s_k ~ ,
\label{comri}
\ee
where the left invariant Maurer--Cartan forms of $SU(2)$,
defined in (\ref{si}), have been used.
The complex structures for the left sector can be similarly
written down
\be
F^-_i = 2 d\rho \wedge \tilde \s_i -  \e_{ijk}
\tilde \s_j \wedge \tilde \s_k ~ ,
\label{comle}
\ee
where $\tilde \s_i$ 
are the right invariant Maurer--Cartan forms of $SU(2)$.
Their explicit expressions can be obtained from (\ref{si})
by letting $(\phi,\th,\psi)\to (-\psi,-\th,-\phi)$,
up to an overall minus sign.
We can readily see that (\ref{comri}),(\ref{comle}) are 
of the general form (\ref{FLR}).

Under $SU(2)$ transformations the variable $\r$ is inert. The 
non--Abelian dual of (\ref{su2}) with respect to a 
vector $SU(2)$ was found in \cite{grnonab}, and we will not
repeat all the steps of the derivation here. 
We only mention that a proper 
unitary gauge choice is $\phi =\psi=0$ among the variables of the $SU(2)$
group element and $v_3=0$ among the Lagrange multipliers. The latter 
choice becomes necessary because, according to our discussion
after (\ref{equmot}), there is a non--trivial 
isotropy group in this case.
After we make the shift $v_2 \to v_2 - \th$ the 
classical solutions for the gauge fields 
$A_\pm = {i\ov 2} \vec A_\pm\cdot \vec \s_{Pauli}$
are 
\ba
 \vec A_\pm & = & {\mp 1\ov 2 v_1^2 \sin^2{\th\ov 2} }\Bigg( 
v_1^2 \del_\pm v_1 + v_1(\sin \th - \th + v_2)\del_\pm v_2, ~
v_1 (\sin\th -\th +v_2) \del_\pm v_1 
\nonumber \\ 
& + & \Big(4 \sin^4{\th\ov 2} +
(\sin\th-\th + v_2)^2 \Big)\del_\pm v_2,~ \pm 2 v_1 \sin^2{\th\ov 2}
\del_\pm v_2 \Bigg) ~ ,
\label{amsu2}
\ea
and the background fields of the dual model are found to be
\ba
ds^2 & = & d\r^2 + d\th^2 + { 1\ov v_1^2 \sin^2{\th \ov 2}} \Bigg (
4 \sin^4{\th \ov 2} dv_2^2 +\Big( v_1 dv_1 + 
(v_2 -\th +\sin\th)dv_2\Big)^2 \Bigg ) ~ ,
\nonumber \\ 
\Phi & = & \ln(v_1^2 \sin^2{\th\ov 2}) ~ ,
\label{dusu2}
\ea
with zero antisymmetric tensor. 
Note that, even though the torsion vanishes, the Ricci tensor
is not zero due to the presence of a non--trivial dilaton. This means
that the manifold is not hyper--Kahler, as the latter property implies 
Ricci flatness \cite{ALFR}. The reason for this apparent paradox 
is of course the fact that 
the original local $N=4$ world--sheet supersymmetry is 
realized in the dual model (\ref{dusu2}) non--locally, 
except for the $N=1$ part.
In the right sector the expressions for the non--local
complex structures are given by  
\be
\tilde F^+_i = C_{ji}(h_-)\Big(
2 d\rho \wedge \cL_j -  \e_{jkl} \cL_k \wedge \cL_l \Big )~ ,
\label{comrinl}
\ee
where $\cL_i = \cL^i{}_\m d X^\m $, $X^\m = \{ \th,v_1,v_2 \}$ and
\ba
(\cL^i{}_\m) =\left( \begin{array} {ccc} 
0 & - 1 & - {v_2-\th\ov v_1} \\
1 & 0 & 0 \\
0 & - \cot {\th\ov 2} & - {2 + \cot{\th\ov 2} (v_2 -\th)\ov v_1} \\
\end{array}
\right) ~ .
\label{lim}
\ea
In the left sector the non--local complex structures
are
\be
\tilde F^-_i = C_{ji}(h_+)\Big(
2 d\rho \wedge \cR_j -  \e_{jkl} \cR_k \wedge \cR_l \Big )~ ,
\label{comlri}
\ee
where $\cR_i = \cR^i{}_\m d X^\m $ and
\ba
(\cR^i{}_\m) =\left( \begin{array} {ccc} 
0 & - 1 &- {v_2-\th \ov v_1} \\
1 & 0 & 0 \\
0 &  \cot {\th\ov 2} & 
 {2 + \cot{\th\ov 2} (v_2 -\th)\ov v_1} \\
\end{array}
\right) ~ .
\label{lirr} 
\ea
The group elements $h_\pm \in SU(2)$ are given by the path 
ordered Wilson lines (\ref{hphm}), with gauge fields (\ref{amsu2}).

\section{ Discussion and concluding remarks}

In this paper we examined the behavior of supersymmetry under
non--Abelian T--duality. We considered
models that are invariant under the 
left action of a general semi--simple group. We gave the general
form of the corresponding $\s$--models as well as of the 
complex structures, in cases that admit extended world--sheet 
supersymmetry, and found their transformation rules under 
non--Abelian duality by utilizing a canonical transformation.
Although, as a general rule,
$N=1$ world--sheet supersymmetry is preserved under duality,
whenever the action of the group on the
complex structures is non--trivial, 
extended supersymmetry seems to be incompatible with non--Abelian
duality. However, this is only an artifact of the description in terms 
of an effective field theory, since non--local world--sheet effects 
restore supersymmetry at the string level.
As examples, $SO(3)$--invariant hyper--Kahler metrics which include the 
Eguchi--Hanson, the Taub--NUT and the Atiyah--Hitchin metrics were 
considered in detail. 
Explicit expressions for the three complex structures were 
given which should be useful in moduli problems in monopole
physics. 
We have also considered 
WZW models and their non--Abelian duals
with respect to the vector action of a subgroup. 
The canonical equivalence of these models was shown by 
explicitly demonstrating that the
algebra obeyed by the Poisson brackets of chiral currents 
of the WZW model is preserved under 
the non--Abelian duality transformation. 
The effect of non--Abelian duality is that the currents 
are represented in terms 
coset parafermions. The latter are non--local
and have non--trivial
braiding properties due to Wilson lines attached to them. 
We believe that this type of canonical equivalence 
is not restricted to just WZW models and their duals but to other
models with vector action of the isometry group.

Non--Abelian duality destroys manifest target space supersymmetry as
well, in the sense that the standard Killing spinor equations 
do not have a solution. 
In fact, 
the breaking of manifest target space
supersymmetry
occurs hand--in--hand with the breaking of local $N=4$ extended
world--sheet supersymmetry. This is attributed to the
relation between Killing spinors and
complex structures \cite{kilcom},
using $F_{\mu\nu} = \bar \xi \G_{\mu\nu} \xi $.
The situation is similar to the case of Abelian duality
\cite{bakasII,BKO,hassand,AAB,sferesto} with the difference that 
the non--local Killing spinors arising after duality do not
define a local $N=2$ world--sheet supersymmetry using
the above relation between Killing spinors and complex structures.
The lowest order effective field
theory is not enough at all to understand the fate of target space
supersymmetry under non--Abelian duality, 
since one has to generate the whole
supersymmetry algebra and not just its truncated part
corresponding to the Killing spinor equations.
In the realization of the 
supersymmetry algebra after non--Abelian
duality massive string modes play a crucial role and a complete
truncation to only the massless modes is inconsistent. 
This becomes apparent
by making contact with the work of Scherk and Schwarz \cite{JSJS}
on coordinate dependent compactifications. The arguments are similar
to the case of Abelian duality and were presented in \cite{kerk95}.

This investigation is part of a program whose goal is to 
use non--trivial stringy effects occuring in
duality symmetries, in 
physical situations that seem paradoxical in
the effective field theory approach. In particular, we would like to
view T--duality as a mechanism of restoring various symmetries,
such as supersymmetry, in a manifest way. An example of how this 
works is based on the background corresponding to 
$SU(2)_k/U(1) \otimes SL(2,\IR)_{-k}/U(1)$. This has $N=4$ world--sheet
supersymmetry which however, is not manifest and is realized using 
parafermions \cite{KAFK}.
An appropriate Abelian duality transformation leads to 
an axionic instanton background with manifest $N=4$ and target space 
supersymmetry restored \cite{sferesto}; 
An equivalent model where target space
supersymmetry was restored by making a moduli parameter dynamical
was considered in \cite{AAB}.
In order to advance these ideas and use non--Abelian duality
as the symmetry restoration mechanism one has to relax 
the condition that an isometry group exists at all, 
since in any case this
is being destroyed by non--Abelian duality.
The notion of non--Abelian duality in the absence of isometries is
now well defined and under the name 
Poisson--Lie T--duality \cite{KliSevI} and the closely related 
quasi--axial--vector duality which was
initiated in \cite{BSthree},
explicitly constructed in \cite{KirObdual} whereas 
its relation to the Poisson-Lie T--duality
was investigated in \cite{TyuBSKO}. 
The idea is to search in various backgrounds of interest 
in black hole Physics or cosmology, 
for ``non--commutative conservation laws'' that
generalize \cite{KliSevI} the usual conservation laws. The hope 
is that in the dual description various properties,
which were hidden, become manifest and possibly resolve certain
paradoxes with field theoretical origin. 
We hope to report work in this direction in the future.

\bs

\centerline{\bf Acknowledgments }
  
\no
I would like to thank I. Bakas for useful discussions and a 
plesant collaboration on related issues.
This work was carried out with the financial support 
of the European Union Research Program 
``Training and Mobility of Researchers'' and is part of the project
``String Duality Symmetries and theories with 
space--time Supersymmetry''.

%%%%%%%%%%%%%%%%%%%%%%%%%%%%%%%%%%%%%%%%%%%%%%%%%%%%%%%%%%%

\appendix
\section{ Derivation of Poisson brackets }
\setcounter{equation}{0}

In this appendix we derive the Poisson brackets of section 3 in a more 
systematic way. Because of the gauging procedure it turns out that 
we are dealing with constrained Hamiltonian systems. 
A consistent way of implementing the constraints was
provided by Dirac (see, for instance, \cite{dirac}).
For our purposes the relevant part of his analysis 
is that given a set of 
second class constraints $\{\vphi_a\}$ one first computes the matrix 
generated by their Poisson brackets 
\be
D_{ab}= \{\vphi_a, \vphi_b\} ~ .
\label{cab}
\ee
In this and in similar computations we are free to use the 
constraints only after calculating their Poisson brackets.
When $D_{ab}$ is invertible one simply postulates that the usual 
Poisson brackets are replaced by the so called Dirac brackets
defined as
\be
\{A,B\}_D = \{A,B\} - \{A,\vphi_a\}D^{-1}_{ab} \{\vphi_b,B\} ~ ,
\label{ABD}
\ee
for any two phase space variables $A$ and $B$.
Then the 
constraints can be strongly set to zero since they have vanishing 
Dirac brackets among themselves and with anything else.

As a very elementary application of this method consider an 
arbitrary action that is first order in time derivatives
\be
S = \int dt A_a(X) \dot X^a ~ .
\label{sasa}
\ee
The conjugate momentum to $X^a$ is given by $P_a=A_a$ and therefore
we cannot solve for the velocity $\dot X^a$ in terms of the momentum
$P_a$.
Hence we impose the constraint 
\be
\vphi_a = P_a - A_a\approx 0 
\label{consstt}
\ee
and follow Dirac's procedure. Using the basic 
Poisson bracket $\{X^a,P_b\}=\d^a{}_b$ we find that the 
matrix (\ref{cab}) is given by
$D_{ab}\equiv M_{ab}= \del_a A_b - \del_b A_a$. Assuming that 
it is invertible and after using (\ref{ABD}) we obtain that
for the general phase space variables $A,B$ the corresponding 
Dirac bracket is given by
\be
\{A,B\}_D = {\del A \ov \del X^a} M^{-1}_{ab} {\del B \ov \del X^b}
~ , ~~~~~~ M_{ab}= {\del A_b\ov \del X^a} - {\del A_a\ov \del X^b}~ .
\label{abbd}
\ee 
This Dirac bracket coincides with the Poisson bracket
postulated in \cite{Wwzw} for the action (\ref{sasa}).
In practice we read off the matrix $M_{ab}$
by simply considering the variation of (\ref{sasa}) 
\be
\d S = \int dt M_{ab} \d X^a \dot X^b ~ .
\label{varsasa}
\ee
In the rest of this appendix as well as in the bulk of the paper
we will call the Dirac bracket (\ref{abbd}) simply a Poisson 
bracket in order to comply with standard terminology in the
literature.

The models we encountered in section 3 belong to the general 
type (\ref{sasa}) where $\s^+$ is considered as the time 
variable, whereas
$\s^-$ is treated as a continuous index.
In that respect our treatment differs from the one in \cite{bowcock}
where $\tau=\s^+ + \s^-$ was taken as the time variable and 
computation of brackets of parafermions was not considered. 

\vskip .1 true cm
\no
\underline{Gauged WZW Models}: The purpose is to reproduce
(\ref{poipar}) in a straightforward way compared to that in 
\cite{BCR,BCH} and
mainly to be able to compare 
with the analogous derivation of 
(\ref{curpar}),(\ref{psipa}),(\ref{siaahj}) which will follow.

Using the definitions (\ref{fhv}) we can write the gauged WZW action as
\be
S= I_{wzw}(h f) - I_{wzw}(h)~ .
\label{sWZW}
\ee
A general variation of the action gives 
\ba
\d S & = & {k\ov \pi} \int 
h^{-1} \d h \Big(
f \del_- (f^{-1} \del_+ f) f^{-1} + f \del_- (f^{-1} h^{-1} \del_+
h f) f^{-1} - \del_- (h^{-1}\del_+ h) \Big) 
\nonumber \\
& & ~~~ +
f^{-1} \d f \Big( \del_- (f^{-1} \del_+ f) 
+ \del_-(f^{-1} h^{-1} \del_+h  f) \Big) ~ .
\label{ds}
\ea
Using for notational convenience the definition
\be
Z_I=\Big( Tr(t^a h^{-1}\d h), Tr(t^A f^{-1} \d f)\Big) ~ ,
\label{ZI}
\ee
we compute the basic Poisson brackets (cf. footnote 7)
%\be
%\Big\{  
%\left(\begin{array}{c}
%h^{-1}\d h \\ f^{-1}\d f \end{array} \right)
%\left(h^{-1}\d h,f^{-1}\d f\right) 
%  \Big\} = -{\pi\ov 2 k}
%M^{-1}(x,y) \e(x-y)~ , 
%\label{po1} 
%\ee
\be
\{Z_I(x),Z_J(y)\} = -{\pi\ov 2 k} M^{-1}_{IJ}(x,y) \e(x-y)~ , 
\label{po1} 
\ee
where the matrix $M(x,y)$ is defined as
\be
M(x,y)= \left( \begin{array} {cc} 
 C_{ab}\big(f(x)f^{-1}(y)\big) - \d_{ab} & C_{aB}(f(x)) \\
C_{bA}(f(y)) & \d_{AB} \end{array} \right)~ .
\label{Mxy}
\ee
Inverting the above matrix and explicitly writing out (\ref{po1})
we obtain
\ba
\{ Tr(t^A f^{-1} \d f)(x),Tr(t^B f^{-1} \d f)(y) \} & = &
{\pi\ov 2k} \e(x-y) \Big( C_{cA}(f(y)) C_{cB}(f(x)) 
- \d_{AB}\Big) ~ ,
\nonumber \\
\{ Tr(t^a h^{-1} \d h)(x),Tr(t^b h^{-1} \d h)(y) \} & = &
{\pi\ov 2k} \e(x-y) \d_{ab}  ~ ,
\nonumber \\
\{ Tr(t^a h^{-1} \d h)(x),Tr(t^B f^{-1} \d f)(y) \} & = &
-{\pi\ov 2k} \e(x-y) C_{aB}(f(x)) ~ .
\label{po2}
\ea
We would like to compute Poisson brackets of the gauged invariant 
quantities $\Psi = {i k\ov \pi} \del_- f f^{-1}$, obeying 
$\del_+ \Psi =0 $ on shell, and $H=-{ik\ov \pi} \del_- h h^{-1}$.
Using the variations  
\ba\d \Psi_A & = & {ik\ov \pi} C_{AB}(f)
Tr\Big(t^B \del_-(f^{-1} \d f)\Big) ~ ,\nonumber \\
\d H_a & = & -{ik\ov \pi} C_{ab}(h)
Tr\Big(t^b \del_-(h^{-1} \d h)\Big) ~ ,
\label{varr}
\ea
 and (\ref{po2}) we obtain
\ba
\{ \Psi_\a(x),\Psi_\b(y)\} & = & 
-{k\ov \pi} \d_{\a\b} \d^{\prime}(x-y) - \Big( f_{\a\b \g} \Psi_\g(y)
+f_{\a\b c} \Psi_c (y)\Big) \d (x-y) 
\nonumber \\
&& -{\pi\ov 2k} f_{c\a\g} f_{c\b\d} \Psi_\g(x) \Psi_\d(y) \e(x-y) ~ ,
\label{pp1} \\
\{ \Psi_a(x),\Psi_b(y)\} & = & f_{abc} \Psi_c(y) \d(x-y) 
-{\pi\ov 2 k} f_{cad} f_{cbe} \Psi_d(x) \Psi_e(y) \e(x-y) ~ ,
\label{pp2} \\
\{ \Psi_a(x),\Psi_\b(y)\} & = & 
-{\pi\ov 2 k} f_{cad} f_{c\b\g} \Psi_d(x) \Psi_\g(y) \e(x-y) ~ ,
\label{pp3}
\ea
and
\ba
\{H_a(x),H_b(y)\} & = & {k\ov \pi} \d_{ab} \d^\prime(x-y) 
- f_{abc} H_c(y) \d(x-y) ~ , \label{pp4} \\
\{H_a(x),\Psi_b(y)\} & = & {k\ov \pi} C_{ab}(h(x)) \d^\prime(x-y) ~ ,
\label{pp5} \\
\{H_a(x),\Psi_\b(y)\} & = & 0 ~ . 
\label{pp6}
\ea 

The form of the action (\ref{sWZW}) suggests that
the equation of motion corresponding to the gauge field $A_+$
has to be imposed as a constraint, i.e.
$\vphi_1^a(x)=\Psi^a(x) \approx 0 $. 
Then, the on shell condition
$F_{+-}=0$ implies the constraint $\vphi_2^a = H^a \approx 0 $
(or $h\approx 1$).
However, due to (\ref{pp4}) and 
(\ref{pp5}) we observe that these cannot be imposed strongly. 
They are second class constraints and the matrix (\ref{cab}),
in the basis
$\vphi_a(x)=\{\vphi_1^a(x),\vphi_2^a(x)\}$, is given by
\be
D_{ab}(x,y) = {k\ov \pi}
\left( \begin{array} {cc}
0 & \d_{ab} \\
\d_{ab} & \d_{ab} \end{array} \right) \d^\prime(x-y) ~ ,
\label{ccaab}
\ee
whereas its inverse is
\be
D^{-1}_{ab}(x,y) = {\pi\ov 2k}
\left( \begin{array} {cc}
-\d_{ab} & \d_{ab} \\
\d_{ab} & 0 \end{array} \right) \e(x-y) ~ .
\label{cab1}
\ee
Then using (\ref{ABD}) we can compute the Dirac brackets of the $\Psi_\a$'s.
It turns out that, 
$\{\Psi_\a,\Psi_\b\}_D \approx \{\Psi_\a,\Psi_\b\}$, hence obtaining 
the result (\ref{poipar}).

\vskip .1 true cm
\no
\underline{Non--abelian duals of WZW Models}: In this case the starting
point is the action (\ref{sss}). Its general variation is given by
\ba
\d S & = & {k\ov \pi} \int 
h^{-1} \d h \Big(
f \del_- (f^{-1} \del_+ f) f^{-1} + f \del_- (f^{-1} h^{-1} \del_+
h f) f^{-1} -\l \del_- (\l^{-1}h^{-1}\del_+ h \l) \l^{-1} 
\nonumber \\
& & - \l \del_-(\l^{-1} \del_+ \l )\l^{-1} \Big) 
+
f^{-1} \d f \Big( \del_- (f^{-1} \del_+ f) 
+ \del_-(f^{-1} h^{-1} \del_+h  f) \Big) 
\nonumber \\
& &- \l^{-1}\d \l \del_-(\l^{-1} h^{-1} \del_+ h \l)~ .
\label{dss}
\ea
Similarly to (\ref{ZI}) we define
\be
Z_I=\Big( Tr(t^a h^{-1}\d h), Tr(t^a\l^a\d \l),
Tr(t^A f^{-1} \d f)\Big) ~ .
\label{ZII}
\ee
These obey (\ref{po1}) with the matrix $M(x,y)$ now defined as
\be
M(x,y)= \left( \begin{array} {ccc} 
C_{ab}\big(f(x)f^{-1}(y)\big) - C_{ab}\big(\l(x)\l^{-1}(y)\big)
&- C_{ab}(\l(x)) & C_{aB}(f(x)) \\
- C_{ba}(\l(y)) & 0 & 0 \\
C_{bA}(f(y)) & 0 & \d_{AB} \end{array} \right)~ .
\label{Mxyy}
\ee
Inverting this matrix we find that the non--zero basic 
Poisson brackets are given by
\ba
\{Tr(t^A f^{-1} \d f)(x),Tr(t^B f^{-1}\d f)(y)\}
& = & - {\pi\ov 2k} \d_{AB} \e(x-y) ~ ,
\nonumber \\
\{Tr(t^a \l^{-1} \d \l)(x),Tr(t^b \l^{-1}\d \l)(y)\}
& = & - {\pi\ov 2k} \d_{ab} \e(x-y) ~ ,
\nonumber \\
\{Tr(t^a \l^{-1} \d \l)(x),Tr(t^B f^{-1}\d f)(y)\}
& = & - {\pi\ov 2k} C_{aB}(\l^{-1}(x)f(x)) \e(x-y) ~ ,
\nonumber \\
\{Tr(t^a h^{-1} \d h)(x),Tr(t^b \l^{-1}\d \l)(y)\}
& = &  {\pi\ov 2k} C_{ab}(\l(y)) \e(x-y) ~ .
\label{many}
\ea
Using them and the variations (\ref{varr}),(\ref{dja})
we calculate the Poisson brackets 
\ba
\{\Psi_A(x),\Psi_B(y)\} & = & -{k\ov \pi} \d_{AB} \d^\prime(x-y) -
f_{ABC} \Psi_C(y) \d(x-y)~ , \label{a1}  \\
\{J_a(x),J_b(y)\} & = & -{k\ov \pi} \d_{ab} \d^\prime(x-y) -
f_{abc} J_c(y) \d(x-y) ~ ,\label{a2}  \\
\{J_a(x),\Psi_b(y)\} & = & -{k\ov \pi} \d_{ab} \d^\prime(x-y) -
f_{abc} J_c(y) \d(x-y)~ , \label{a3}  \\
\{J_a(x),\Psi_\b(y)\} & = & 0 ~ , \label{a4}  
\ea
and
\ba
\{H_a(x),J_b(y)\} & = & - {k\ov \pi} C_{ab}(h(x)) \d^\prime(x-y)
- f_{dbc} C_{ad}(h(y)) J_c(y) \d(x-y)~ , \label{a5} \\
\{H_a(x),H_b(y)\} & = & \{H_a(x),\Psi_A(y)\} = 0 ~ .
\label{a6}
\ea
As in the case of gauged WZW models we have to impose 
the equation of motion
corresponding to $A_+$ as a constraint, i.e.,
$\vphi_1^a(x)=\Psi^a(x)-J^a(x)\approx 0$, 
as well as $\vphi_2^a(x)=H^a(x)\approx 0 $ 
corresponding to $F_{+-}=0$. 
Since they cannot imposed strongly 
we again follow Dirac's procedure. 
We first compute the matrix (\ref{cab})
\be
D_{ab}(x,y) = {k\ov \pi}
\left( \begin{array} {cc}
0 & C_{ab}(\l(x) \l^{-1}(y)) \\
C_{ba}(\l(x) \l^{-1}(y)) & 0 \end{array} \right) \d^\prime(x-y) ~ ,
\label{ccab1}
\ee
and its inverse 
\be
D^{-1}_{ab}(x,y) = {\pi\ov 2k}
\left( \begin{array} {cc}
0 & C_{ab}(\l(y)\l^{-1}(x)) \\
C_{ba}(\l(y)\l^{-1}(x)) & 0 \end{array} \right) \e(x-y) ~ .
\label{caba1}
\ee 
Then using (\ref{ABD}) we 
obtain that the Dirac bracket $\{\Psi_A,\Psi_B\}_D$ coincides
with the corresponding Poisson bracket (\ref{a1}).
As a consistency check the Dirac brackets of the $J_a$'s 
should coincide with the Dirac brackets of the $\Psi_a$'s
because the constraint $\vphi^a_1$ is imposed strongly. 
This can be verified using (\ref{ABD}) 
and the explicit 
form of the matrix $D^{-1}_{ab}$ in (\ref{caba1}).
We note that this is not the case for the corresponding
Poisson brackets as one can see from (\ref{a3}),(\ref{a4}).

Finally, let us mention that the conclusion we have 
reached about WZW models and their non--Abelian duals,
would have of course been the same even if we had worked,
within Dirac's general framework,
with the action (\ref{sssf}) instead of (\ref{sss}).

%%%%%%%%%%%%%%%%%%%%%%%%%%%%%%%%%%%%%%%%%%%%%%%%%%%

%.............

\end{document}